\documentclass[
reprint,
superscriptaddress,
amsmath,amssymb,
aps,
prl,
longbibliography]{revtex4-2}

\usepackage{xspace}
\usepackage{graphicx}
\usepackage{bm}
\usepackage[colorlinks=true,urlcolor=blue,linkcolor=blue,citecolor=blue,bookmarks=false]{hyperref}
\usepackage{amsmath}
\usepackage{amssymb}
\usepackage{bbold}
\usepackage{soul}
\usepackage{dsfont}
\usepackage{xcolor}

\newcounter{myequation}
\makeatletter
\@addtoreset{equation}{myequation}
\makeatother

\newcounter{myfigure}
\makeatletter
\@addtoreset{figure}{myfigure}
\makeatother

\newcommand{\customref}[2]{\hyperref[#1]{\ref*{#1}#2}}

\definecolor{Ured}{HTML}{cc0000}
\definecolor{Ublue}{HTML}{1f65cf}
\definecolor{Ugreen}{HTML}{198a11}

\newcommand{\ve}[1]{\mathbf{#1}}

\DeclareMathOperator{\Tr}{Tr}

\newcommand{\ie}[0]{i.e.\@\xspace}

\newfont{\tensy}{cmsy10}

\renewcommand{\S}[0]{\hat{\mathcal{H}}}

\newcommand{\banvec}[1]{\hat{\mathbf{a}}^{\vphantom\dagger}_{#1}}
\newcommand{\bcrvec}[1]{\hat{\mathbf{a}}^{\dagger}_{#1}}

\newcommand{\im}{\mathrm{i}}

\newcommand{\NN}[1]{\langle #1 \rangle}

\newcommand{\absolute}[1]{\left| #1 \right|}
\newcommand{\expv}[1]{\left\langle #1 \right\rangle}

\newcommand{\spin}[1]{\hat{\mathbf{S}}_{#1}}
\newcommand{\spinz}[1]{\hat{S}^{z}_{#1}}

\newcommand{\spinc}[2]{\hat{S}^{#2}_{#1}}

\newcommand{\omegac}{\omega_\mathrm{c}}

\begin{document}
\title{Dissipation-induced order: the $S=1/2$ quantum spin chain coupled to an ohmic bath}

\author{Manuel Weber}
\affiliation{Max Planck Institute for the Physics of Complex Systems, N\"othnitzer Str.~38, 01187 Dresden, Germany}

\author{David J. Luitz}
\affiliation{Physikalisches Institut, Universit\"at Bonn, Nussallee 12, 53115 Bonn, Germany}
\affiliation{Max Planck Institute for the Physics of Complex Systems, N\"othnitzer Str.~38, 01187 Dresden, Germany}

\author{Fakher F. Assaad}
\affiliation{$\mbox{Institut f\"ur Theoretische Physik und Astrophysik, Universit\"at W\"urzburg, 97074 W\"urzburg, Germany}$}
\affiliation{\mbox{W\"urzburg-Dresden Cluster of Excellence ct.qmat, Am Hubland, 97074 W\"urzburg, Germany}}

\date{\today}

\begin{abstract}
We consider an $S=1/2$ antiferromagnetic quantum Heisenberg chain where each site is coupled to an independent bosonic bath with ohmic dissipation. The coupling to the bath preserves the global SO(3) spin symmetry. Using large-scale, approximation-free quantum Monte Carlo simulations, we show that any finite coupling to the bath suffices to stabilize long-range antiferromagnetic order. This is in stark contrast to the isolated Heisenberg chain where spontaneous  breaking of the SO(3) symmetry is forbidden by the Mermin-Wagner theorem. A linear spin-wave theory analysis  confirms that the memory of the bath and the concomitant retarded interaction stabilize the order. For the Heisenberg chain, the ohmic bath is a marginal perturbation so that exponentially large system sizes are  required to observe long-range order at small couplings. Below this length scale, our numerics is dominated by a crossover regime where spin correlations show different power-law behaviors in space and time. We discuss the experimental relevance of this crossover phenomena. 
\end{abstract}

\maketitle

\textit{Introduction.---}%
Real quantum systems are seldom isolated \cite{Weiss_book,breuer_theory_2002}.  The natural question to ask is if the coupling to the environment will 
trigger new  phenomena,  and, if so,  at  which  energy- or  timescale. This  question  is not only  relevant in the realm  of quantum simulation or computing where decoherence is  a limiting factor \cite{preskill_quantum_2018}, but  also  in the solid state.
A prominent example for this are experiments on KCuF$_3$ \cite{Lake05}, a quasi-one-dimensional material with weak interchain coupling. In this material, surrounding chains can be viewed as a weakly-coupled environment modifying the behavior of the chain: At  \textit{high} energies,   neutron-scattering  experiments are  remarkably well reproduced  by the   two-spinon continuum of the isolated Heisenberg model; at \textit{low}  energies,  the  environment  dominates, leading to the binding of spinons into spin waves.

One of our motivations is to understand the physics of chains of magnetic adatoms deposited on a metallic substrate \cite{Toskovic16}. Starting from an effective description of the magnetic adatoms in terms of a one-dimensional $S=1/2$ Heisenberg chain
with a Kondo-type coupling to the substrate \cite{Danu19,Danu20,Danu22},
one can use Hertz-Millis theory \cite{PhysRevB.14.1165,PhysRevB.48.7183} to integrate out the bath and obtain in second-order perturbation theory a retarded interaction in space and time between the spin degrees of freedom. This interaction is governed by the spin susceptibility of the two-dimensional electron gas,
$\chi_0(i-j,\tau-\tau')$, where $i$ and $j$  denote the positions of the magnetic adatoms and $\tau,\tau'$ are points in imaginary time; it has a different decay in space (quartic) and time (quadratic). 
In our modeling, we will neglect the spatial decay since it is irrelevant at the Heisenberg critical point
\footnote{Power counting  shows  that  interactions decaying  with a   
quartic power law are  irrelevant  at the  Heisenberg  critical point \cite{Cardy96_Book,Goldenfeld}}
and focus on the effect of retardation of the interaction in (imaginary) time  \cite{PhysRevLett.79.4629,PhysRevLett.97.076401,PhysRevB.86.035455,friedman2019dissipative,2015EL....10957001D}. 
This allows us to simplify the model further and instead of a metallic substrate we introduce independent ohmic baths described by noninteracting bosons as in the celebrated Caldeira-Leggett model \cite{Caldeira81}, leading to the same retarded interaction in time if the bath is integrated out.

Spin chains in the presence of dissipation have been considered 
in the absence of the Berry phase
within an  $\epsilon$ expansion \cite{Pankov04} as well as with classical Monte Carlo  methods  \cite{Werner05}. 
Simulations were based on a lattice discretization of the nonlinear sigma model \cite{Werner05}, but without the
topological $\theta$ term that is relevant for half-integer spin chains and renders them critical \cite{Haldane83}.
In the absence of the Berry phase, these   spin  models   account for  massive  phases, so that
a finite  coupling  to the bath is   required  to  trigger a 
phase  transition
from a disordered phase at weak coupling to 
an ordered phase at strong coupling.  This  breakdown  of the  Mermin-Wagner theorem \cite{mermin_absence_1966,hohenberg_existence_1967} stems  from the   fact   that   the  ohmic bath 
induces long-ranged  retarded interactions.  
Calculations for  the   quantum  XXZ  chain    with site  ohmic 
dissipation   coupling  to the $z$ component of the spin  were  carried out in Ref.~\cite{PhysRevLett.113.260403}.


 
In this Letter, we focus on the SO(3)-symmetric quantum Heisenberg chain with spin-symmetric coupling to the ohmic baths.
In  contrast  to previous  work  \cite{Werner05}, 
we directly solve the $S=1/2$ quantum spin problem; this automatically  takes  into  account  the Berry  phase so  that  the isolated spin chain 
becomes  critical.
The recently introduced  wormhole  algorithm \cite{weber2021quantum} permits positive-sign quantum Monte Carlo simulations
for very large system sizes, which allows us to systematically study the approach toward the thermodynamic limit.
We find that the  coupling to the ohmic  bath is marginal and that \textit{any} coupling strength modifies the
low-energy physics of the $S=1/2$ chain, stabilizing long-range antiferromagnetic order. Our results reveal a nontrivial finite length scale, separating suppressed correlations at short distances from the emergence of order at long distances. We expect that this length scale is observable in experiments for finite spin chains.

\textit{Model.---}%
We consider the one-dimensional $S=1/2$ antiferromagnetic Heisenberg model
\begin{align}
\label{eq:ham_s}
\hat{H}_\mathrm{s}
	=
	J \sum_i \spin{i} \cdot \spin{i+1} \, ,
\end{align}
where we use the exchange coupling $J=1$ as the unit of energy.  
 Its ground state shows critical antiferromagnetic correlations given for long distances by 
 $C(r)= \langle \spinz{0} \spinz{r} \rangle \propto (-1)^r \left( \ln r \right)^{1/2} r^{-1}$, where the power law is tied to the global SO(3) spin symmetry \cite{affleck_exact_1998}.

To study the effects of dissipation on the Heisenberg chain, we introduce an independent bosonic bath coupled to each spin component $\hat S_i^\alpha$. The total Hamiltonian is given by 
$\hat{H} = \hat{H}_\mathrm{s} + \hat{H}_\mathrm{sb}$, with 
\begin{align}
\label{eq:ham_sb}
\hat{H}_\mathrm{sb}
	=
	\sum_{iq} \omega_q \, \bcrvec{iq} \cdot \banvec{iq}
	+ \sum_{iq} \lambda_q  \, \big( \bcrvec{iq} + \banvec{iq} \big)  \cdot \spin{i} \, .
\end{align}
Here, $\bcrvec{iq}$, $\banvec{iq}$ are 3-component vectors of bosonic creation and annihilation operators.   
The bath consists of a continuum of modes $q$ with frequency $\omega_q$ and spin-boson
coupling $\lambda_q$.  
Our model satisfies the global SO(3) rotational symmetry generated by the
total angular momentum 
$  \hat{\ve{J}}_{\text{tot}}   =  \sum_{iq}      \hat{\ve{Q}}_{iq}   \times   \hat{\ve{P}}_{iq}      +  \sum_{i}  \hat{\ve{S}}_i$,
where $\hat{\ve{Q}}_{iq} = \frac{1}{\sqrt{2}} ( \bcrvec{iq} +    \banvec{iq} ) $ and
 $\hat{\ve{P}}_{iq} = \frac{\im}{\sqrt{2}} ( \bcrvec{iq}  - \banvec{iq} ) $
 are the bosonic position and momentum operators, respectively.   
The effects of the bath on the spin system are fully determined by the
spectral density
$
J(\omega)
	=
	\pi \sum_{q} \lambda_{q}^2 \, \delta(\omega - \omega_{q})
$.
An ohmic bath corresponds to  a power-law spectrum
\begin{align}
\label{eq:spectrum}
J(\omega)
	=
	2\pi \alpha \, J^{1-s} \omega^s \, ,
	\qquad 0< \omega < \omegac \, ,
\end{align}
with exponent $s=1$ \cite{Weiss_book}. 
Here, we introduced the dimensionless coupling constant $\alpha$ and the frequency cutoff $\omega_\mathrm{c}$.

The bath can be integrated out exactly and the partition function
$Z = Z_\mathrm{b} \Tr_\mathrm{s} \hat{\mathcal{T}}_\tau \, e^{-\S}$
is fully determined by the spin subsystem
$\S = \S_\mathrm{s} + \S_\mathrm{ret}$.
The spin-boson coupling in Eq.~(\ref{eq:ham_sb}) leads to a retarded spin-spin interaction
\begin{align}
\label{eq:Hret}
\S_\mathrm{ret}
	=
	- 
	\iint_0^\beta d\tau d\tau'  \sum_i
	K(\tau-\tau') \,
	\spin{i}(\tau) \cdot \spin{i}(\tau') \, ,
\end{align}
which encodes the memory of the bath. It is mediated by the bath propagator 
\begin{align}
K(\tau) 
	= 
	 \int_0^{\omegac} d\omega \,
	\frac{J(\omega)}{\pi}
	\frac{\cosh[\omega (\beta/2 - \tau)]}{2\sinh[\omega \beta /2]} \, ,
\end{align}
where
$0 \leq \tau < \beta$ and $K(\tau+\beta) = K(\tau)$. 
Here, $\beta=1/T$ is the inverse temperature.
The power-law spectrum in Eq.~(\ref{eq:spectrum}) yields $K(\tau) \sim 1/\tau^{1+s}$ for $\omegac \tau \gg 1$.

The retarded interaction can invalidate the Mermin-Wagner theorem  and  produce long-range order even in one  spatial dimension. In the Supplemental Material \cite{SM}, we provide a linear spin-wave theory analysis of  our  model, which shows that at   large $S$ spin waves  do not  destabilize antiferromagnetic long-range  order in the presence of dissipation.
Further insight comes  from considering the  isolated  spin chain,  
and,  at this  critical  point, computing the  scaling dimension of the    retarded  interaction.
One  obtains
$ \Delta =  1-s $   such  that the  ohmic  case, $s=1$,  is  marginal  \cite{Cardy96_Book,Goldenfeld}.   The  goal  here is   to  investigate numerically 
 if  the coupling  is marginally  relevant or  irrelevant.

\textit{Method.---}%
For our simulations, we used an exact quantum Monte Carlo method for
retarded interactions \cite{PhysRevLett.119.097401} that samples a
diagrammatic expansion of $Z / Z_\mathrm{b}$ in $\S_\mathrm{s} + \S_\mathrm{ret}$.
Our approach is based on the stochastic series expansion \cite{PhysRevB.43.5950}
with global directed-loop updates \cite{PhysRevE.66.046701} and makes use
of efficient wormhole moves \cite{weber2021quantum} recently developed
for retarded spin-flip interactions as in Eq.~(\ref{eq:Hret}). The time dependence of
$K(\tau-\tau')$ only enters during the diagonal updates and is sampled exactly
using inverse transform sampling \cite{weber2021quantum}; we set
$\omegac / J =10$, similar to Ref.~\cite{PhysRevLett.113.260403}.
At $\alpha=0$, Lorentz invariance guarantees convergence to the ground state at inverse temperatures $\beta \propto L$. This is no longer true for $\alpha>0$, so that we ensure convergence in temperature for all results, as demonstrated in the Supplemental Material \cite{SM}.
For the largest system sizes we reach $\beta J \approx 10{,}000$, and we use
periodic boundary conditions.
For a detailed description of our method see Ref.~\cite{weber2021quantum}.

\begin{figure}[t]
  \includegraphics[width=0.95\linewidth]{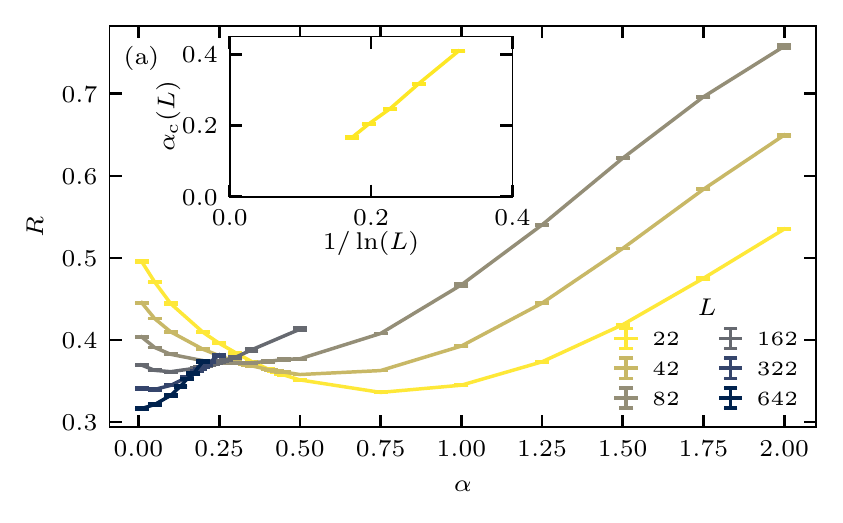} \\
  \includegraphics[width=0.95\linewidth]{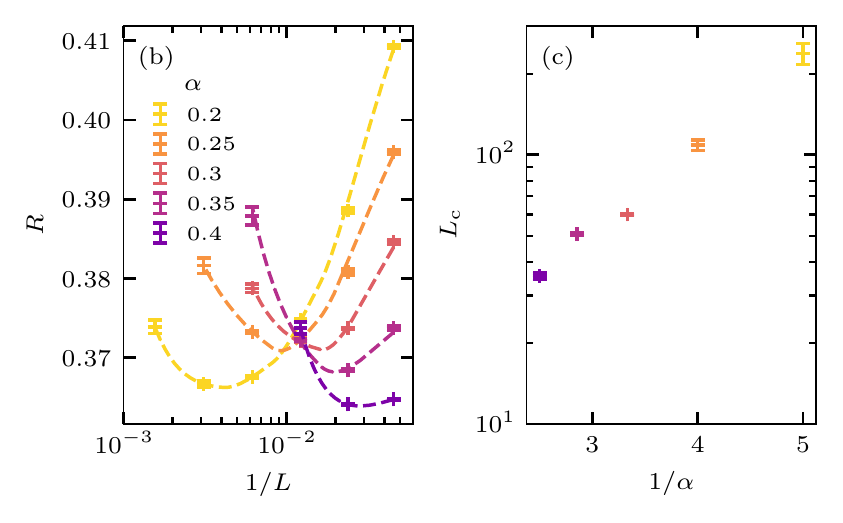}
  \caption{\label{fig:afm-crossings}%
(a)~Antiferromagnetic correlation ratio $R$ as a function of the spin-boson coupling $\alpha$
for different system sizes $L$. A finite-size scaling
of the crossings $\alpha_\mathrm{c}(L)$ between data pairs $\{L,2L-2\}$ is shown as an inset. 
(b)~$R(1/L)$ for different $\alpha$.
The minima of $R(1/L)$ define a crossover scale $L_\mathrm{c}(\alpha)$ beyond which $R$ increases.
$L_\mathrm{c}$ is estimated using spline fits and shown in (c).
The crossover scale is consistent with an exponential scaling $L_\mathrm{c} \propto \exp(\zeta / \alpha)$.
  }
\end{figure}

\textit{Results.---}%
To probe for long-range order, we compute the equal-time spin structure factor  defined as 
\begin{align}
S(q)
= \frac{1}{L} \sum_{ij} e^{\im q (i-j)} \big\langle {\spinc{i}{z} \spinc{j}{z}} \big\rangle \, . 
\end{align} 
Owing to the spin-rotational symmetry of our model,
it is sufficient to consider only the $z$ component of the spin.
We also calculate the correlation ratio \cite{doi:10.1063/1.3518900,Kaul15},
\begin{align}
R
	=
	1 - \frac{S(Q+\delta q)}{S(Q)} \, ,
\end{align}
at the ordering momentum $Q=\pi$ and with resolution
$\delta q=2\pi/L$, as it is particularly useful to detect quantum phase transitions.  $R$   captures  $\left( \xi/L \right)^2$   where  $\xi$ is  the  correlation length \cite{Caracciolo93}.
It scales to unity (zero) in the ordered (disordered) phase and corresponds to a renormalization-group-invariant quantity at criticality.
Figure~\ref{fig:afm-crossings}(a)  shows  temperature-converged  results of $R$ for each chain length $L$ and  coupling strength $\alpha$.   
For  \textit{large}  values of $\alpha$,   the correlation ratio grows, thus lending support to  long-range  antiferromagnetic  order  as   suggested by linear  spin-wave theory  \cite{SM}.   To  understand the limit of strong bath coupling $\alpha$,  we consider  $J=0$ in Eq.~\eqref{eq:ham_s}.     In this case,  $  \hat{\ve{J}}_{i,\text{tot}}   =  \sum_{q}      \hat{\ve{Q}}_{iq}   \times   \hat{\ve{P}}_{iq}      +   \hat{\ve{S}}_i$   is a good  quantum number  such  that  the  ground state    for  each    site, consisting  of  a  spin and the   bath, has  a  half-integer   angular  momentum and  is  hence  at least   twofold  degenerate, \ie, the  bath cannot screen  the spin  degree of freedom. This leads  to a macroscopic  degeneracy,  which is  lifted at finite $J$   by  the  
onset  of long-ranged order,  as  shown in Fig.~\ref{fig:afm-crossings}(a).

We now   turn  our   attention to the weak-coupling limit.  
 Considering pairs of chain lengths,  we  observe  that  the  crossing of $R(\alpha,L)$ and $R(\alpha, 2L-2)$  at  $\alpha_\mathrm{c}(L)$   systematically   drifts to lower  values of $\alpha$.   As apparent  from the inset  of   Fig.~\ref{fig:afm-crossings}(a)  and for  our  considered lattice  sizes,  $ \alpha_\mathrm{c}(L) \simeq 1/\ln(L)$.   Figure~\ref{fig:afm-crossings}(b)  shows the correlation ratio $R$ at fixed   coupling  $\alpha$ and  as a function of lattice  size, revealing a characteristic  length  scale $L_\mathrm{c}$  at  which $R$ shows a  minimum.  The  $\alpha$  dependence of 
$L_\mathrm{c}$, shown in Fig.~\ref{fig:afm-crossings}(c),   is consistent  with an  exponential  law,  $L_\mathrm{c}(\alpha)  \propto   e^{\zeta/\alpha} $,    
suggesting that the coupling to the ohmic bath is marginally relevant.
As a  consequence, exponentially large lattices are required to observe ordering  in the regime of small $\alpha$.

\begin{figure}[t]
    \centering
    \includegraphics[width=\columnwidth]{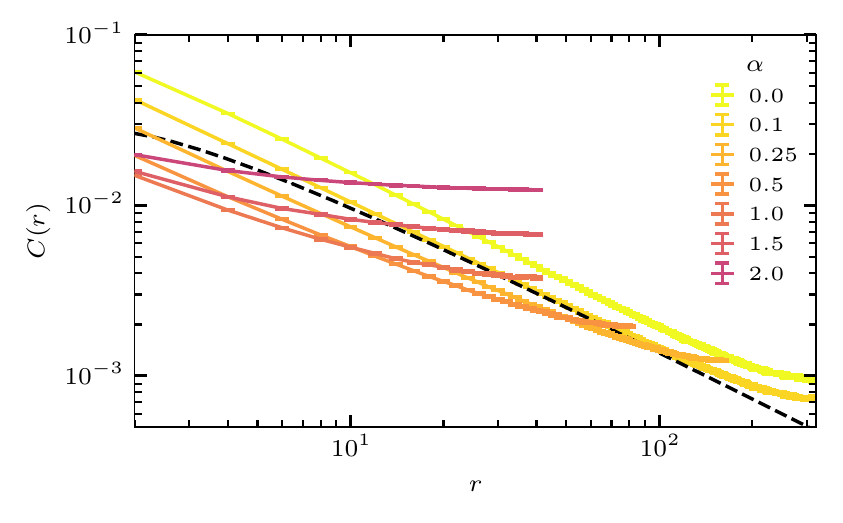}
    \caption{Real-space correlation function $C(r)$ for even $r$ at the largest available $L\in\{82,162,322,642\}$ for different $\alpha$.
    For $\alpha=0$, the exact asymptotic form (dashed line),
    $C(r) \sim (-1)^r (\ln r)^{1/2} / [(2\pi)^{3/2} r]$, is approached very slowly \cite{affleck_exact_1998}.
    A finite-size analysis of the boundary effects near $r = L/2$ can be found in the Supplemental Material \cite{SM}.
    }
    \label{fig:realspace}
\end{figure}

The length scale $L_\mathrm{c}$,  beyond  which the correlation ratio $R$ grows,  is  revealed   by the  real-space correlations $C(r)$   shown in Fig.~\ref{fig:realspace}.  At $\alpha=0.1$,  this length  scale lies beyond the  lattice  sizes accessible in our simulations,  and  $C(r)$ is, up to an overall scaling  factor,  not   distinguishable  from the correlations in the Heisenberg  model.  
We  interpret the   renormalization 
of the  short-ranged spin-spin correlations  in terms of  entanglement  between  bath and spin   degrees of freedom.  At $\alpha = 0.25$,    the   correlation ratio grows   for  $L \gtrsim 82$,
as can be seen in Fig.~\ref{fig:afm-crossings}(b).   The  length  scale $L_\mathrm{c}$  marks   a   distinct  departure   from  the   Heisenberg  scaling and  a leveling  off  of the  spin-spin correlations in Fig.~\ref{fig:realspace}.      Ultimately   for $\alpha \geq 1 $,     $L_\mathrm{c}$  drops  below our smallest  system size,  the Heisenberg scaling  is not apparent any more, and the data  clearly support long-ranged order.      We  also note  that   while  initially  decreasing,   the   magnitude of  the    short-ranged spin-spin correlations grows for   \textit{large}  values  of  $\alpha$. 

\begin{figure}[b]
  \includegraphics[width=\linewidth]{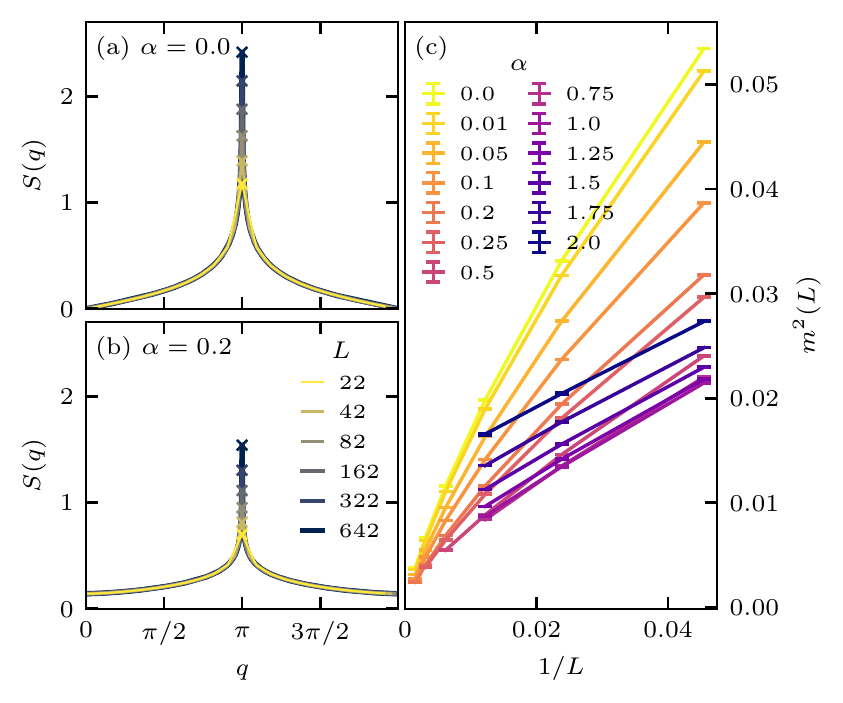}
  \caption{\label{fig:orderparam}%
  Finite-size dependence of the spin structure factor $S(q)$ for (a) $\alpha=0.0$ and (b) $\alpha=0.2$.
(c) Finite-size scaling of the order parameter $m^2(L) = S(q=\pi)/L$ for different $\alpha$.  
  }
\end{figure}
Figure~\ref{fig:orderparam}  displays the   spin structure  factor $S(q)$  as   well as  the  square of the antiferromagnetic order parameter,
\begin{equation}
  m^2(L)   =   \frac{1}{L}   S(q = \pi) \, . 
\end{equation} 
In the  absence of  the bath,   $S(q = \pi)$  diverges logarithmically [Fig.~\ref{fig:orderparam}(a)]  so that $ m^2(L \rightarrow \infty )$   vanishes [Fig.~\ref{fig:orderparam}(c)]. 
 For  \textit{ large } bath couplings $\alpha$,  we  observe a finite order parameter $m^2(L \rightarrow  \infty ) > 0 $ in Fig.~\ref{fig:orderparam}(c), in accordance  with  our  analysis of the correlation ratio.  
 At small $\alpha$, distinguishing  $m^2(L \rightarrow  \infty ) $  from zero  becomes challenging.    In this  limit,    the  data  of  Fig.~\ref{fig:realspace}    show  that   the 
 spin-spin correlations   decay as  $1/r$ for  $r<L_c $  before  leveling off.    Hence, we  conjecture  that
$\lim_{L \rightarrow \infty} m^2(L)   \propto   1/L_\mathrm{c}(\alpha) \propto e^{-\zeta/\alpha} $.
The  structure  factor    equally  reveals  the  value  of the  total  spin via 
$S(q=0) =  \frac{1}{3L} \big\langle \big(\sum_{i} \hat{\ve{S}}_i\big)^2 \big\rangle$.
For the Heisenberg  chain,   the  total  spin  is  a good quantum number  and  vanishes at  zero  temperature  on  any  finite lattice [Fig.~\ref{fig:orderparam}(a)],
whereas any nonzero coupling to the  bath breaks  this  symmetry [Fig.~\ref{fig:orderparam}(b)].   The finite  value of $S(q=0) $    reflects the  entanglement  of the spin  chain and the bath.

\begin{figure}
    \centering
    \includegraphics[width=\columnwidth]{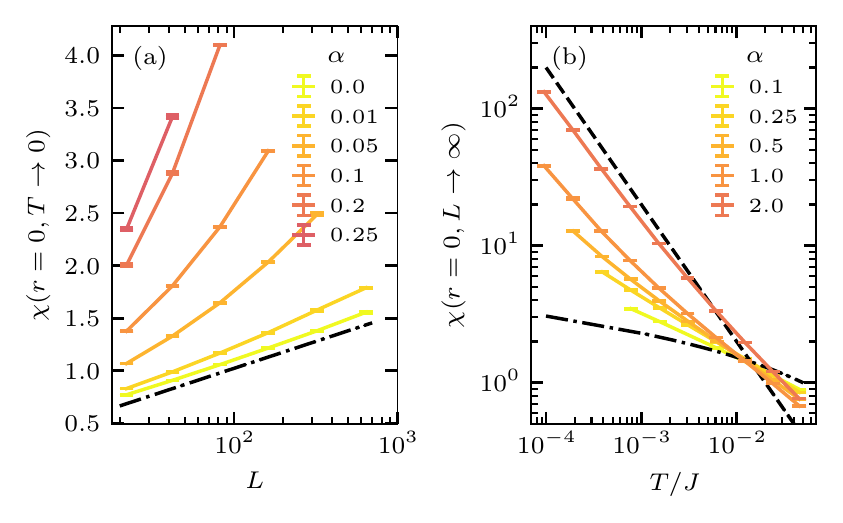}
    \caption{Local spin susceptibility $\chi(r=0)$ (a) for $T\to 0$ as a function of $L$
    and (b) for $L\to \infty$ as a function of $T$. The dot-dashed lines indicate
    an $\ln L$ dependence in (a) and an $\ln \beta$ dependence in (b), whereas the dashed line in (b) corresponds to
    the $\chi \sim 1/T$ behavior expected for the ordered state.
    }
    \label{fig:sus}
\end{figure}

From  the  equal-time  correlation  functions,   one  would conclude that   in the  small  $\alpha$ limit  and   at  distances   smaller  than $L_\mathrm{c}$   one 
observes  the  physics of the Heisenberg model.   This  turns out not to be the case.    One of the   defining properties of the Heisenberg  chain is Lorentz  invariance that   renders space and  time  interchangeable.    In  Fig.~\ref{fig:sus}   we   show the  local  spin  susceptibility 
\begin{equation}
	\chi(r=0, L,\beta)   =  \frac{1}{L} \sum_{i=1}^{L} \int_{0}^{\beta}  d \tau  \,   \langle  \hat{S}^{z}_{i} (\tau) \hat{S}^{z}_{i} (0)  \rangle   
\end{equation}
where $\hat{S}^{z}_{i} (\tau) =  e^{\tau \hat{H}} \hat{S}^{z}_{i}   e^{-\tau \hat{H}} $.   As   detailed in the Supplemental Material  \cite{SM},   for the Heisenberg  model  
$\chi(r=0, L  ,\beta \rightarrow \infty) \propto \ln(L) $  and $\chi(r=0, L\rightarrow \infty  ,\beta ) \propto \ln( \beta) $.    
In  Fig.~\ref{fig:sus}(a)   this scaling 
is confirmed for the Heisenberg model.  Of   particular interest is the dataset  at $\alpha=0.1$.   Here, our   lattice sizes are  smaller than  $L_\mathrm{c}(\alpha) $  and  the  real-space  correlations $C(r)$ in Fig. \ref{fig:realspace} are not   distinguishable from those  of the Heisenberg model.    However, $\chi(r=0) $  shows  marked   deviations from  the  logarithmic scaling of  the Heisenberg  model.   Hence, in the  crossover  regime  where  our  system sizes are smaller than $L_\mathrm{c}(\alpha) $,  the  local  susceptibility  is  not  controlled by the  Heisenberg  fixed point.   In  particular,    our  data are  consistent   with correlations in time that  decay  slower than $1/\tau$. 
For  larger   values of $\alpha$ our  system  sizes  exceed  $L_\mathrm{c}$  such that  we  can pick  up  long-range order in the local susceptibility, \ie,
 $\chi(r=0, L\rightarrow \infty  ,\beta ) \propto \beta $.   As apparent from Fig.~\ref{fig:sus}(b),  we  observe this  behavior  for  large values of  $\alpha$.  
 Note that for any value of $\alpha$  we  expect  the local    susceptibility  to  reveal  long-range order   for  lattice  sizes  $L> L_\mathrm{c}(\alpha) $. 

\textit{Discussion.---}%
Our results  demonstrate  the efficiency of our quantum Monte Carlo method  for  retarded interactions.  Unprecedentedly   large lattices at  very low  
temperatures  can  be reached, necessary to reveal the physics of  dissipative  $S=1/2$  quantum  spin chains. 

To  best  interpret  our   results,  it is convenient to  consider  our  model in the $\alpha$ versus $s$ plane.   For  $s < 1$ ($s>1$)  the  coupling to the  bath  is  relevant  (irrelevant).   For  $s > 1$  we conjecture   that there will  be a phase  transition  between the  Heisenberg chain and a phase  with long-ranged order at finite value of $\alpha_\mathrm{c}(s)$.       We note that for the 1+1 dimensional  nonlinear Ising and O(2)  sigma models,  such a   dissipation-induced ordering  transition has  been studied \cite{Werner05}.    At  $s=1$ (considered  here),  the coupling to the bath is marginal,  and  our  results  are   consistent   with the  interpretation that it is \textit{marginally  relevant}.  As a  consequence,  we  observe a  very   slow  flow:  in the small $\alpha$   regime    lattice  sizes   greater  than   $L_\mathrm{c}(\alpha)  \propto e^{\zeta/\alpha} $  are  required  to reveal  long-range  ordering.    The  physics  in the crossover  regime  $ L < L_\mathrm{c}(\alpha)  $ is  particularly interesting. Here, the real-space  correlation functions    decay  as  $1/r$  akin to the  Heisenberg  chain.  On the other  hand,  the  imaginary-time  correlations  reveal  a  breakdown of  Lorentz  invariance  and fall off  much slower  than $1/\tau$.      A possible  interpretation  is the proximity  to  the  quantum phase  transition  at $\alpha_\mathrm{c}(s) $  for  $s > 1$.  As seen  in  Ref.~\cite{Werner05}    for the 
1+1 dimensional  nonlinear O($n$)  sigma models,  such transitions   have dynamical  exponents  
$z>1$
with  $\tau^{-1/z} $   decay in imaginary time.    Such an interpretation of the data  can be tested   since  the  phase  diagram  of our model in the $\alpha$--$s$ plane    can be  investigated  with our quantum Monte Carlo  algorithm.  

Our model is relevant for the understanding  of chains of magnetic adatoms on   two-dimensional    metallic surfaces \cite{Toskovic16}.  These   experiments  
are    typically    limited   to  a small  number of adatoms.  The  fact that  the  coupling  to the  bath is marginally  relevant implies   that the physics 
of these  chains  will   be captured  by the  crossover  regime.   Furthermore, spin-orbit  coupling,  generically  present at surfaces, will break 
 the SO(3)  spin symmetry down to SO(2).    Similar calculations as presented here but for the  XXZ  chain  are hence of  particular  interest.   

The finite-temperature   and  dynamical properties of  our model   will   reveal   how  the two-spinon  continuum   will evolve  when  coupled to the bath.  While  the high-energy features of the    dynamical  spin structure  factor   will reveal the two-spinon continuum,  the  low-energy  features   should   be
captured   by the spin-wave  theory  \cite{SM} of  damped magnons  with spectral weight emerging above  $\omega \propto k^2$.

\begin{acknowledgments}
We thank   B.~Danu,  T.~Grover  and  M.~Vojta   for many illuminating  discussions  on  related  research. 
 F.~F.~A. thanks the DFG for funding via  the W\"urzburg-Dresden
  Cluster of Excellence on Complexity and Topology in Quantum Matter ct.qmat
  (EXC 2147, Project No. 390858490). D.~J.~L.~acknowledges support by the DFG through SFB 1143 (Project No.~247310070) and the cluster of excellence ML4Q (EXC2004, Project No. 390534769).
\end{acknowledgments}


%

\clearpage

\newcommand{\Aan}[1]{\hat{A}^{\vphantom\dagger}_{#1}}
\newcommand{\Acr}[1]{\hat{A}^{\dagger}_{#1}}
\newcommand{\Ban}[1]{\hat{B}^{\vphantom\dagger}_{#1}}
\newcommand{\Bcr}[1]{\hat{B}^{\dagger}_{#1}}

\newcommand{\Acohan}[1]{A_{#1}}
\newcommand{\Acohcr}[1]{\bar{A}_{#1}}
\newcommand{\Bcohan}[1]{B_{#1}}
\newcommand{\Bcohcr}[1]{\bar{B}_{#1}}

\newcommand{\Alcohan}[1]{{\alpha}_{#1}}
\newcommand{\Alcohcr}[1]{\bar{\alpha}_{#1}}
\newcommand{\Becohan}[1]{{\beta}_{#1}}
\newcommand{\Becohcr}[1]{\bar{\beta}_{#1}}

\newcommand{\kvec}{\mathbf{k}}
\newcommand{\rvec}[1]{\mathbf{r}_{#1}}
\newcommand{\del}{\boldsymbol{\delta}}

\renewcommand{\S}[0]{\mathcal{S}}


\stepcounter{myequation}
\stepcounter{myfigure}

\renewcommand{\thefigure}{S\arabic{figure}}
\renewcommand{\thesection}{S\arabic{section}}
\renewcommand{\thetable}{S\arabic{table}}
\renewcommand{\theequation}{S\arabic{equation}}

\onecolumngrid

%
%
%



\centerline{\bf\large Supplemental Material} \vskip3mm
\centerline{\bf\large for} \vskip3mm
\centerline{\bf\large Dissipation-induced order: the $S=1/2$ quantum spin chain coupled to an ohmic bath} \vskip3mm

\section{Finite-temperature analysis of the order parameter}

\begin{figure}[b]
    \centering
    \includegraphics[width=\columnwidth]{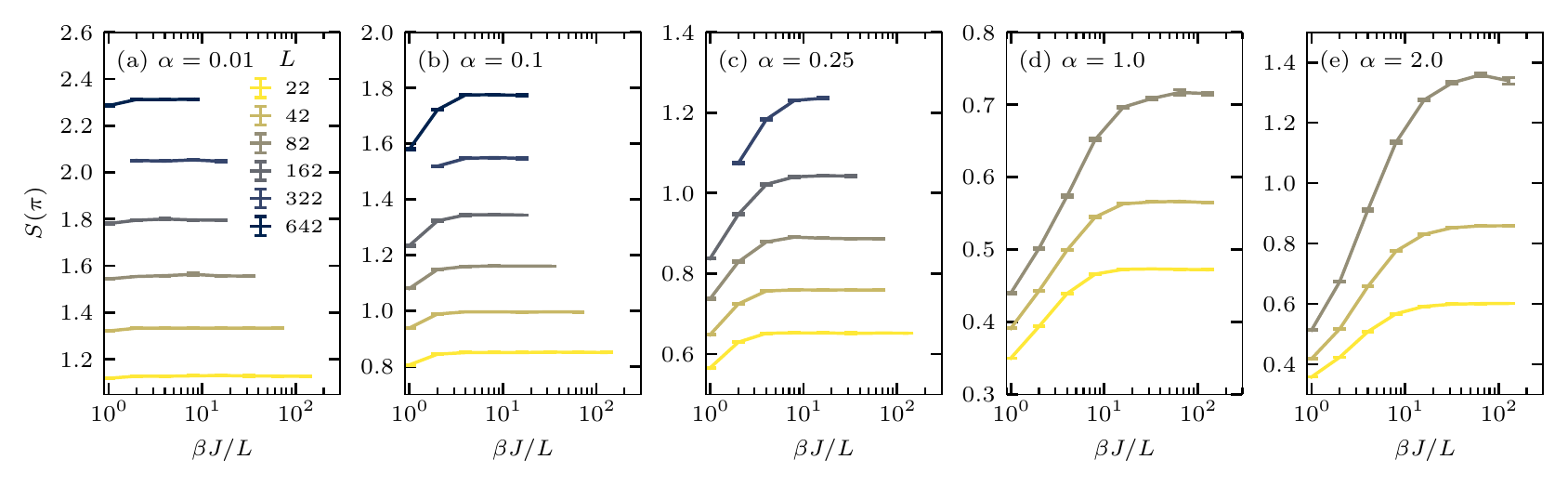}
    \includegraphics[width=\columnwidth]{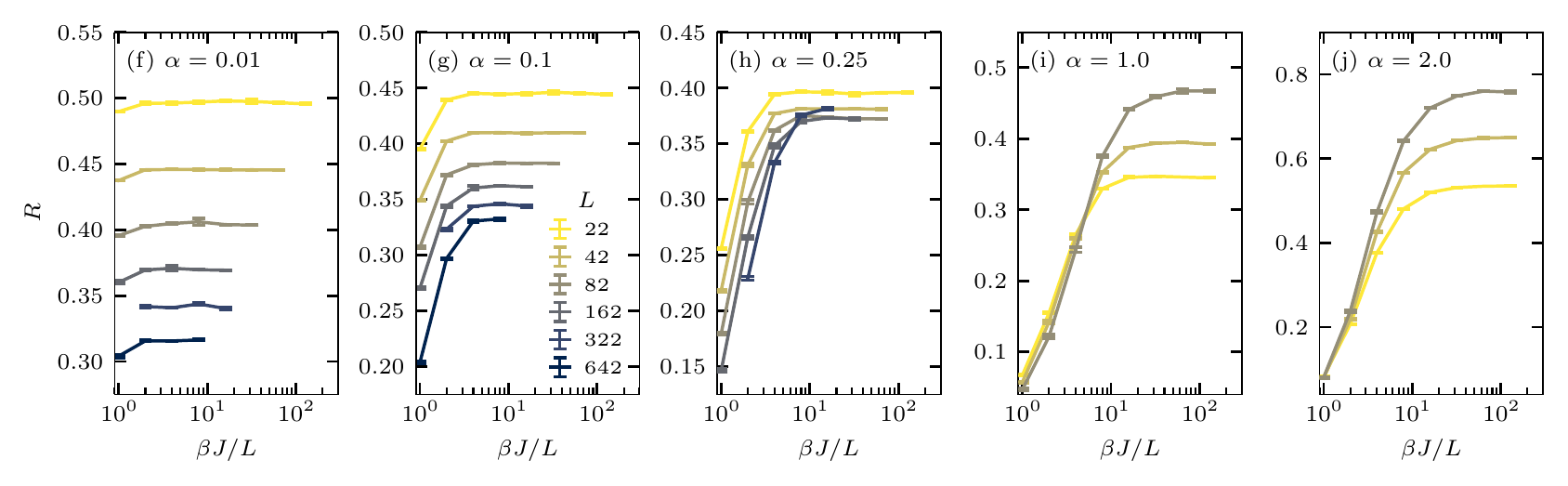}
    \caption{Convergence of (a)--(e) the order parameter $S(\pi)$ and (f)--(j) the
    correlation ratio $R$ as a function of $\beta / L$ for different spin-boson couplings $\alpha$ and lattice sizes $L$.
    We have chosen all data points according to $\beta / L = 2^n$, $n\in\mathds{N}_0$.
    }
    \label{figsm:fs_order}
\end{figure}

Figure~\ref{figsm:fs_order} shows the inverse-temperature dependence of the order parameter $S(\pi)$
and the correlation ratio $R$ for different spin-boson couplings $\alpha$ and different system sizes $L$.
To determine the ground-state properties of the dissipative Heisenberg chain, we have to make
sure that our observables are converged for each parameter set. For $\alpha = 0$, it is sufficient
to choose inverse temperatures $\beta = 2L$ to get converged results and we reach
system sizes up to $L=642$.
Here, the $\beta \sim L^z$ scaling with dynamical exponent $z=1$ is a consequence
of conformal invariance.
The coupling to the bath breaks conformal invariance so that we have to simulate at increasingly lower temperatures
with increasing $\alpha$ and $L$. For the strongest couplings $\alpha=1.0$ or $2.0$, we can only reach temperature
convergence up to $L=82$ where $\beta J/L \approx 2^7$ ($\beta J \approx 10{,}000$) is required.
Moreover, when doubling the system sizes in Figs.~\ref{figsm:fs_order}(d), \ref{figsm:fs_order}(e), \ref{figsm:fs_order}(i), and \ref{figsm:fs_order}(j) we approximately need an additional
factor of $2$ in $\beta / L$ for temperature convergence. This is a strong hint towards
$z=2$ scaling,
as predicted by our spin-wave calculation (see below). At smaller couplings, we also
expect a crossover toward $z=2$ scaling, but system sizes are still too small to resolve this.
The temperature dependence of $S(\pi)$ and $R$ in Fig.~\ref{figsm:fs_order} is representative
for other equal-time observables.

\section{Finite-size dependence of the real-space correlations}

\begin{figure}[h]
    \centering
    \includegraphics[width=\columnwidth]{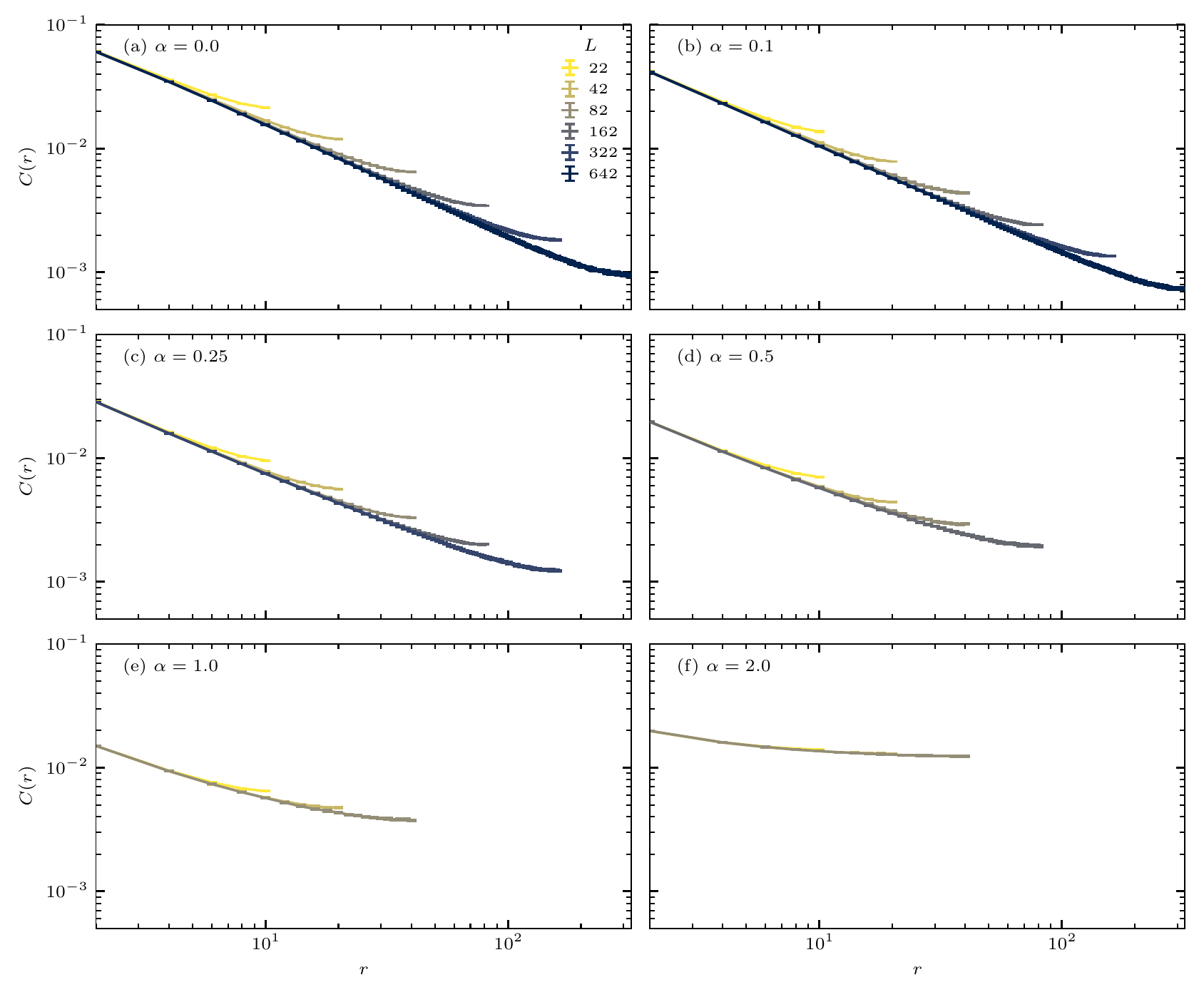}
    \caption{Finite-size dependence of the real-space spin-spin correlation function $C(r)$ for different
    spin-boson couplings $\alpha$. For each lattice size $L$, $C(r)$ has converged to the zero-temperature
    result according to Fig.~\ref{figsm:fs_order}.
    }
    \label{figsm:fs_realspace}
\end{figure}

Figure~\ref{figsm:fs_realspace} shows a finite-size analysis of the real-space spin-spin correlation function $C(r)$
for different couplings $\alpha$. In the absence of conformal invariance, we cannot use the conformal distance
to get rid of boundary effects. Therefore, we have to analyze $C(r)$ for different $L$ to estimate the maximum
distance $r$ that has already converged to the $L\to \infty$ limit. For a detailed discussion of $C(r)$ we refer
to the main part of our paper.

\newpage

\section{Linear spin-wave theory for the dissipative Heisenberg antiferromagnet}

In the following, we perform the linear spin-wave approximation for the antiferromagnetic Heisenberg model coupled to a bosonic bath.
We consider the Hamiltonian $\hat{H} = \hat{H}_\mathrm{s} + \hat{H}_\mathrm{sb}$ with
\begin{gather}
\label{eq:ham_s}
\hat{H}_\mathrm{s}
	=
	J \sum_{\NN{ij}} \spin{i} \cdot \spin{j} \, ,
\\
\label{eq:ham_sb}
\hat{H}_\mathrm{sb}
	=
	\sum_{iq} \omega_q \, \bcrvec{iq} \cdot \banvec{iq}
	+ \sum_{iq} \lambda_q  \, \big( \bcrvec{iq} + \banvec{iq} \big)  \cdot \spin{i} \, .
\end{gather}

\subsection{Holstein-Primakoff transformation}

First, we use the Holstein-Primakoff transformation to represent the spins in terms of bosonic operators.
We assume that the model is defined on a bipartite lattice.
On sublattice $\mathrm{A}$, we expand around the spin-$S$ state in $z$ direction, \ie,
\begin{align}
i\in \text{sublattice A:} \qquad
\spinz{i} &= S - \Acr{i} \Aan{i} \, , \\
\spinc{i}{+} &= \sqrt{2S - \Acr{i} \Aan{i}} \, \Aan{i} = \sqrt{2S} \, \Aan{i} + \mathcal{O}(S^{-1/2}) \, , \\
\spinc{i}{-} &= \Acr{i} \sqrt{2S - \Acr{i} \Aan{i}} = \sqrt{2S} \, \Acr{i} + \mathcal{O}(S^{-1/2}) \, ,
\end{align}
whereas on sublattice $\mathrm{B}$ we expand around $-S$, \ie,
\begin{align}
j\in \text{sublattice B:} \qquad
\spinz{j} &= \Bcr{j} \Ban{j} - S \, , \\
\spinc{j}{+} &= \Bcr{j} \sqrt{2S - \Bcr{j} \Ban{j}} = \sqrt{2S} \, \Bcr{j} + \mathcal{O}(S^{-1/2}) \, , \\
\spinc{j}{-} &= \sqrt{2S - \Bcr{j} \Ban{j}} \, \Ban{j} = \sqrt{2S} \, \Ban{j} + \mathcal{O}(S^{-1/2}) \, .
\end{align}
Note that we have approximated the spin flip operators 
in such a way that our final result will be correct up to $\mathcal{O}(S^{0})$ corrections.
For the contribution of the system, we obtain
the familiar result for the antiferromagnetic Heisenberg model,
\begin{align}
\label{eq:ham_s_HP}
\hat{H}_\mathrm{s}
	=
	\frac{LqJS^2}{2} +
	JS  \sum_{\NN{i,j}}  \left(  \Aan{i} \, \Ban{j}   +  \Acr{i} \, \Bcr{j} \right)
	+ q J S \sum_{i\in\mathrm{A}} \Acr{i} \, \Aan{i} 
	+ q J S \sum_{j\in\mathrm{B}} \Bcr{j} \, \Ban{j}
	+ \mathcal{O}(S^{0})
		 \, ,
\end{align}
where $q$ is the coordination number of the lattice. Note that we drop all $\mathcal{O}(S^{0})$ terms.
The spin-boson part becomes
\begin{align}
\hat{H}_\mathrm{sb}
\nonumber
	=
	\sum_{iq} \omega_q \, \bcrvec{iq} \cdot \banvec{iq}
	&+ \sum_{i\in \mathrm{A}} \sum_q \lambda_q  \, \big( \bcrvec{iq} + \banvec{iq} \big)  \cdot 
	\begin{pmatrix}
	\sqrt{S/2} \, \big[\Acr{i} + \Aan{i} \big] \\
	\im \sqrt{S/2} \, \big[ \Acr{i} - \Aan{i} \big] \\
	S - \Acr{i} \Aan{i}
	\end{pmatrix}
	\\
	&+ \sum_{j\in \mathrm{B}} \sum_q \lambda_q  \, \big( \bcrvec{jq} + \banvec{jq} \big)  \cdot 
	\begin{pmatrix}
	\sqrt{S/2} \, \big[\Ban{j} + \Bcr{j} \big] \\
	\im \sqrt{S/2} \, \big[ \Ban{j} - \Bcr{j} \big] \\
	\Bcr{j} \Ban{j} - S
	\end{pmatrix}
	+ \mathcal{O}(S^{-1/2})
	\, .
\label{eq:ham_sb_HP}
\end{align}
At this stage, it is still important to keep the $\mathcal{O}(S^{0})$ term in the spin-$z$ component
of $\hat{H}_\mathrm{sb}$.
 
\subsection{Integrating out the bosonic bath in the path-integral formulation} 
 
Second, we introduce the coherent-state path integral for the partition function,
\begin{gather}
Z 
	=
	\int \mathcal{D}(\Acohcr{},\Acohan{})
	\int \mathcal{D}(\Bcohcr{},\Bcohan{}) \,
	e^{-\S_\mathrm{s}[\Acohcr{},\Acohan{},\Bcohcr{},\Bcohan{}]}
	\int \mathcal{D}(\bar{\mathbf{a}},\mathbf{a}) \, 
	e^{-\S_{\mathrm{sb}}[\Acohcr{},\Acohan{},\Bcohcr{},\Bcohan{},\bar{\mathbf{a}},\mathbf{a}]} \, ,	
\end{gather}
where we represent all operators in terms of bosonic coherent states.
Here, $\S_\mathrm{s}$ and $\S_\mathrm{sb}$ are the actions that correspond
to Eqs.~(\ref{eq:ham_s_HP}) and (\ref{eq:ham_sb_HP}), respectively.
For details on coherent states and the path-integral formalism, see Ref.~\cite{Negele}.
In a third step, we integrate out the bosonic bath and obtain
\begin{align}
Z 
	=
	Z_\mathrm{b}
	\int \mathcal{D}(\Acohcr{},\Acohan{})
	\int \mathcal{D}(\Bcohcr{},\Bcohan{}) \,
	e^{-\S_\mathrm{s}[\Acohcr{},\Acohan{},\Bcohcr{},\Bcohan{}] - \S_\mathrm{ret}[\Acohcr{},\Acohan{},\Bcohcr{},\Bcohan{}] } \, ,
\end{align}
where $Z_\mathrm{b}$ is the partition function of the noninteracting bath.
The contribution of the system to the action is given by
\begin{align}
\nonumber
\S_\mathrm{s}
	=
	\frac{L \beta qJS^2}{2} 
	&+ JS \int_0^\beta d\tau \sum_{\NN{i,j}}  \left[  \Acohan{i}(\tau) \, \Bcohan{j}(\tau)   +  \Acohcr{i}(\tau) \, \Bcohcr{j}(\tau)  \right] \\
	&+ \int_0^\beta d\tau \left[ 
		\sum_{i\in\mathrm{A}} \Acohcr{i}(\tau) \left( \partial_\tau + q J S \right) \Acohan{i}(\tau) +
		\sum_{j\in\mathrm{B}} \Bcohcr{j}(\tau) \left( \partial_\tau + q J S \right)  \Bcohan{j}(\tau)
		 \right],
\end{align}
where the terms $\bar A_i \partial_\tau A_i$ and $\bar B_j \partial_\tau B_j$  encode the bosonic Berry phase.
Furthermore, we get the retarded interaction,
\begin{align}
\nonumber
\S_\mathrm{ret}
	=
	&- \iint_0^\beta d \tau d\tau' \, K(\tau-\tau')
	\sum_{i\in\mathrm{A}}
	\left\{
	S^2
	- S \left[ \Acohcr{i}(\tau) \, \Acohan{i}(\tau) +  \Acohcr{i}(\tau') \, \Acohan{i}(\tau') \right]
	+ S \left[ \Acohan{i}(\tau) \, \Acohcr{i}(\tau') +  \Acohcr{i}(\tau) \, \Acohan{i}(\tau') \right]
	\right\}
	\\
	&- \iint_0^\beta d \tau d\tau' \, K(\tau-\tau')
	\sum_{j\in\mathrm{B}}
	\left\{
	S^2
	- S \left[ \Bcohcr{j}(\tau) \, \Bcohan{j}(\tau) +  \Bcohcr{j}(\tau') \, \Bcohan{j}(\tau') \right]
	+ S \left[ \Bcohan{j}(\tau) \, \Bcohcr{j}(\tau') +  \Bcohcr{j}(\tau) \, \Bcohan{j}(\tau') \right]
	\right\} 
\end{align}
that stems from integrating out the bosons.
Note that the diagonal spin-boson interaction leads to an equal-time contribution with a
time-dependent correction of $\mathcal{O}(S^0)$ which we omit, whereas the spin-flip terms lead to a nonlocal
interaction in imaginary time. This retarded interaction is mediated by the bath propagator,
\begin{align}
\label{Eq:ktau_def}
K(\tau) 
	= 
	 \int_0^{\omegac} d\omega \,
	\frac{J(\omega)}{\pi}
	\frac{\cosh[\omega (\beta/2 - \tau)]}{2\sinh[\omega \beta /2]} \, ,
\qquad 
\mathrm{where} 
\qquad
J(\omega)
	=
	\begin{cases}
	2\pi \alpha J^{1-s} \, \omega^s    &   0< \omega < \omegac \\
	0 & \mathrm{else}
	\end{cases}
	\, .
\end{align}
We assume a power-law spectrum $J(\omega)$ with exponent $s$ and frequency cutoff $\omega_\mathrm{c}$,
where $s=1$ corresponds to an ohmic bath. We have included the Heisenberg exchange constant $J$ in the definition
of $J(\omega)$ so that the spin-boson coupling $\alpha$ becomes dimensionless. With this definition, $K(\tau)$ does
not change for $\omega_\mathrm{c} \tau \gg 1$ if we change $\omega_\mathrm{c}$.

\subsection{Diagonalization of the action}

In order to diagonalize the action, we define the Fourier transformation of the bosonic fields,
\begin{align}
\Acohan{i}(\tau) = \frac{1}{\sqrt{\beta L'}} \sum_{\kvec n} e^{\im (\Omega_n \tau - \kvec \cdot \rvec{i})} \, \Acohan{\kvec n} \, ,
\qquad\qquad
\Acohan{\kvec n} = \frac{1}{\sqrt{\beta L'}} \int_0^\beta d\tau \sum_{i\in\mathrm{A}} e^{-\im (\Omega_n \tau - \kvec \cdot \rvec{i})} \, \Acohan{i}(\tau) \, ,
\\
\Bcohan{j}(\tau) = \frac{1}{\sqrt{\beta L'}} \sum_{\kvec n} e^{-\im (\Omega_n \tau - \kvec \cdot \rvec{j})} \, \Bcohan{\kvec n} \, ,
\qquad\qquad
\Bcohan{\kvec n} = \frac{1}{\sqrt{\beta L'}} \int_0^\beta d\tau \sum_{j\in\mathrm{B}} e^{\im (\Omega_n \tau - \kvec \cdot \rvec{j})} \, \Bcohan{j}(\tau) \, ,
\end{align}
where we introduced the number of sites per sublattice, $L'=L/2$, the lattice vector $\rvec{i}$, the momentum $\kvec$, and the bosonic Matsubara frequencies $\Omega_n = 2\pi n / \beta$, $n\in \mathds{Z}$.
For the Heisenberg interaction, we obtain
\begin{align}
\S_\mathrm{s}
	&=
	\frac{L\beta qJS^2}{2} +
	 qJS \sum_{\kvec n}  \left[ \gamma_\kvec \, \Acohan{\kvec n} \Bcohan{\kvec n} + \gamma^*_\kvec \, \Acohcr{\kvec n} \Bcohcr{\kvec n} \right]
	 + \sum_{\kvec n} \left[  \left( \im \Omega_n + qJS \right) \Acohcr{\kvec n} \Acohan{\kvec n} + \left( - \im \Omega_n + qJS \right) \Bcohcr{\kvec n} \Bcohan{\kvec n} \right] \, ,
\end{align}
where $\gamma_\kvec = q^{-1} \sum_{\del} e^{\im \kvec \cdot \del}$ only depends on the translation vectors $\del_1, \dots, \del_q$ between nearest-neighbor sites.
For the retarded interaction, we get
\begin{align}
\S_\mathrm{ret}
	=
	- L \beta S^2 K_0
	+ 2S \sum_{\kvec n} \left(K_0 - K_n\right) \left[ \Acohcr{\kvec n} \Acohan{\kvec n} + \Bcohcr{\kvec n} \Bcohan{\kvec n} \right] \, .
\end{align}
Here, we used the Matsubara transformation of the boson propagator,
\begin{align}
\label{Eq:Kn_def}
K_n 
	=
	 \int_0^\beta d \tau \, e^{\im \Omega_n \tau}  K(\tau)
	=
	\int_0^{\omega_\mathrm{c}} d\omega \frac{J(\omega)}{\pi} \frac{\omega}{\omega^2 + \Omega_n^2} \, .
\end{align}
Note that $K_{-n} = K_n$.
Eventually, the full interaction becomes 
$\S_\mathrm{s} + \S_\mathrm{ret}= L \beta S^2 \left( qJ/2 - K_0 \right) + \S_1 + \mathcal{O}(S^0)$, where
\begin{align}
\S_1
	=
	\sum_{\kvec n}
	\begin{pmatrix}
	\Acohcr{\kvec n} & \Bcohan{\kvec n}
	\end{pmatrix}
	\begin{pmatrix}
	\im \Omega_n + qJS + 2S\left(K_0 - K_n \right) & qJS \, \gamma^*_\kvec \\
	qJS \, \gamma_\kvec & -\im \Omega_n + qJS + 2S\left(K_0 - K_n \right)
	\end{pmatrix}
	\begin{pmatrix}
	\Acohan{\kvec n} \\
	\Bcohcr{\kvec n}
	\end{pmatrix}
	\, .
\end{align}
The action $\S_1$ can be diagonalized using the real-valued canonical Bogoliubov transformation
\begin{align}
\begin{pmatrix}
	\Alcohan{\kvec n} \\
	\Becohcr{\kvec n}
\end{pmatrix}
=
\begin{pmatrix}
	u_{\kvec n} & v_{\kvec n} \\
	v_{\kvec n} & u_{\kvec n}
\end{pmatrix}
\begin{pmatrix}
	\Acohan{\kvec n} \\
	\Bcohcr{\kvec n}
\end{pmatrix}
\, ,
\qquad
\begin{pmatrix}
	\Acohan{\kvec n} \\
	\Bcohcr{\kvec n}
\end{pmatrix}
=
\begin{pmatrix}
	u_{\kvec n} & -v_{\kvec n} \\
	-v_{\kvec n} & u_{\kvec n}
\end{pmatrix}
\begin{pmatrix}
	\Alcohan{\kvec n} \\
	\Becohcr{\kvec n}
\end{pmatrix}
\, ,
\end{align}
which fulfills $u_{\kvec n}^2 - v_{\kvec n}^2 = 1$ in order to preserve the measure of the path integral.
We determined the matrix elements to be
\begin{align}
u_{\kvec n}
	=
	\sqrt{ \frac{1}{2} \left( \frac{1}{\sqrt{1-w_{\kvec n}^2}} + 1 \right) } \, ,
\qquad
v_{\kvec n}
	=
	\sqrt{ \frac{1}{2} \left( \frac{1}{\sqrt{1-w_{\kvec n}^2}} - 1 \right) } \, ,
\qquad
w_{\kvec n}
	=
	\frac{qJS \absolute{\gamma_\kvec} }{qJS + 2S\left(K_0 - K_n \right)} \, ,
\end{align}
such that the action takes the diagonal form
\begin{gather}
\S_1
	=
	\sum_{\kvec n}
        \left[  \left(\im \Omega_n + \epsilon_{\kvec n}\right) \Alcohcr{\kvec n} \Alcohan{\kvec n} 
        + \left(-\im \Omega_n + \epsilon_{\kvec n}\right) \Becohcr{\kvec n} \Becohan{\kvec n} \right]
	\, ,
\\
\text{where} \quad
\epsilon_{\kvec n} 
	= 
	\sqrt{ \left[   qJS + 2S\left(K_0 - K_n \right)   \right]^2 - \left( q J S \right)^2 \absolute{\gamma_\kvec}^2 } \, .
\label{Eq:dispersion}
\end{gather}
For $\alpha=0$, we recover the magnon dispersion of the antiferromagnetic Heisenberg model,
$\epsilon_{\kvec} = qJS \sqrt{ 1 - |{\gamma_\kvec}|^2 }$. 
In one dimension, we have $\gamma_k = \cos(k)$ such that
$\epsilon_{k} = 2JS \absolute{\sin(k)}$, \ie, the dispersion is linear at low energies.
For any finite spin-boson coupling $\alpha$, $\epsilon_{kn}$ shows a nontrivial dependence
on the Matsubara frequency and therefore cannot be interpreted as a dispersion relation anymore.
To get access to the spectrum, one has to perform the analytic continuation of the Green's function
which is proportional to $1/(\Omega_n^2 + \epsilon_{\kvec n}^2)$.
Eventually, the spin-boson coupling will lead to a continuous spectral function; this will be discussed
elsewhere.

We want to take a closer look at the frequency dependence of $\epsilon_{\kvec n}$ in Eq.~(\ref{Eq:dispersion}).
To this end, we calculate $K_n$ using the definitions in Eqs.~(\ref{Eq:Kn_def}) and (\ref{Eq:ktau_def}).
We find that $K_0 = 2\alpha J^{1-s} \omega_\mathrm{c}^s / s$ diverges with the cutoff frequency. For the
ohmic case with $s=1$, we can calculate $K_n$ for any $\omega_\mathrm{c}$, \ie,
$K_n = 2 \alpha \left[ \omega_\mathrm{c} - \Omega_n \arctan(\omega_\mathrm{c} / \Omega_n) \right]$.
We find that the divergent term in the cutoff frequency drops out such that the limit $\omega_\mathrm{c} \to \infty$ is
well defined. More generally, we have
\begin{align}
\label{Eq:kernel_limit}
\lim_{\omega_\mathrm{c} \to \infty} \left( K_0 - K_n \right)
	=
	\frac{\pi \alpha J^{1-s} \absolute{\Omega_n}^s}{\sin(\pi s/2)} \, ,
	\qquad 0 < s < 2 \, ,
\end{align}
so that the coupling to the bosonic bath leads to a term proportional to $|\Omega_n|^s$.

\subsection{Stability of the spin-wave solution}

In the following, we want to test whether antiferromagnetic order remains stable within the spin-wave solution.
To this end, we calculate the expectation value of the boson occupation number,
\begin{align}
N_S =
\frac{1}{L\beta} \int_0^\beta d\tau
\left[
 \sum_{i\in\text{A}}  \big\langle {\Acr{i}(\tau) \, \Aan{i}(\tau)} \big\rangle
 + \sum_{j\in\text{B}} \big\langle {\Bcr{j}(\tau) \, \Ban{j}(\tau)} \big\rangle 
 \right] \, .
\end{align}
If $N_S$ diverges, the leading-order fluctuations of the spin-wave solution destroy the antiferromagnetic
ground state, whereas the ordered state remains stable as long as $N_S$ is finite.
After transforming this expectation value into the diagonal basis of our path-integral solution, we get
(note that we omit a constant shift of $1/2$)
\begin{align}
\nonumber
N_S
	&=
	\frac{1}{\beta L} \sum_{\kvec n} 
	\left( u_{\kvec n}^2 + v_{\kvec n}^2 \right)
	\expv{\Alcohcr{\kvec n} \Alcohan{\kvec n} + \Becohcr{\kvec n} \Becohan{\kvec n} }
	=
	\frac{1}{\beta L} \sum_{\kvec n} 
	\left( u_{\kvec n}^2 + v_{\kvec n}^2 \right) 
	\left[ \frac{1}{\im \Omega_n + \epsilon_{\kvec n}} + \frac{1}{-\im \Omega_n + \epsilon_{\kvec n}} \right]
	\\
	&=
	\frac{2}{\beta L} \sum_{\kvec n} 
	\frac{qJS + 2S \left(K_0 - K_n\right)}{\Omega_n^2 + \epsilon_{\kvec n}^2} \, .
\end{align}

Now we can take the continuum limit and use the frequency dependence of $K_n$
given in Eq.~(\ref{Eq:kernel_limit}) to obtain
\begin{align}
N_S
	=
	\frac{4}{(2\pi)^{d+1}} \int_0^{\omega_\mathrm{c}} d\omega 
	\int_{\mathrm{BZ}} d\kvec \,
	\frac{qJS + \tilde{\alpha} \, \omega^s}
	{\omega^2 + \left[ qJS + \tilde{\alpha} \, \omega^s   \right]^2 - \left( q J S \right)^2 \absolute{\gamma_\kvec}^2 } \, ,
\end{align}
where we have gained a factor of $2$ because the integrand is an even function in $\omega$.
Furthermore, we defined $\tilde{\alpha}=2\pi \alpha S J^{1-s}/\sin(\pi s/2)$.
For the one-dimensional case, the integral becomes
\begin{align}
\label{Eq:integral_final_m1}
N_S
	&=
	\frac{1}{\pi^2} \int_0^{\omega_\mathrm{c}} d\omega 
	\int_0^{2\pi} dk \,
	\frac{2JS + \tilde{\alpha} \, \omega^s}
	{\omega^2 + (2JS)^2 \sin^2(k) +  4JS \tilde{\alpha} \, \omega^s + \tilde{\alpha}^2 \omega^{2s} } 
	\\
	&=\frac{2}{\pi}  \int_0^{\omega_\mathrm{c}} d\omega \, 
	\frac{2JS + \tilde{\alpha} \, \omega^s}
	{\sqrt{\left( \omega^2 +  4JS \tilde{\alpha} \, \omega^s + \tilde{\alpha}^2 \omega^{2s} \right)
	\left[(2JS)^2 + \omega^2 +  4JS \tilde{\alpha} \, \omega^s + \tilde{\alpha}^2 \omega^{2s} \right]
	} } 
	\, .
\label{Eq:integral_final}
\end{align}
In the last step, we have used $\int_0^{2\pi} dk \left[a+\sin^2(k)\right]^{-1} = 2\pi /\sqrt{a\left(1+a\right)}$. 

Now we want to determine whether $N_S$ diverges or remains finite. To this end, we analyze the low-frequency limit
of the integrand in Eq.~(\ref{Eq:integral_final}). The leading behavior is determined by 
$1/\sqrt{\omega^2 + c \, \alpha \, \omega^s}$. For $\alpha=0$, it becomes $1/\omega$ and therefore
the integral diverges. However, for any finite coupling to the bath, the leading behavior becomes
$1/\sqrt{\omega^{s}}$ which is integrable for $s<2$.
As a result, the coupling to the bath stabilizes long-range antiferromagnetic order for large $S$.

Finally, from Eq.~(\ref{Eq:integral_final_m1}) we can also deduce that conformal invariance is broken by the coupling to the bath. For
$\omega\ll 1$ and $k\ll 1$, the propagator becomes $1/(\omega^2 + v_\mathrm{s}^2 k^2 + \alpha'  \omega^s)$ where $v_\mathrm{s}$
corresponds to the spin-wave velocity. Hence, we have $z=2$ for the coupling to an ohmic bath with $s=1$.

\section{Scaling of the local susceptibility for the   Heisenberg   chain}

At zero temperature and in the thermodynamic limit, the Lorentz symmetry inherent to the Heisenberg spin chain leads to
\begin{equation}
    \langle \hat{S}^z(r,\tau) \, \hat{S}^z(0,0) \rangle     \propto  \frac{1}{\sqrt{r^2 + (v_\mathrm{s} \tau)^2}} \, .
\end{equation}
Here, $v_\mathrm{s}$  corresponds  to the spin-wave  velocity and we have omitted logarithmic corrections.  If we now consider a finite-sized system of length  $L$, the system will  have a  spin gap   set  by  $\Delta_\mathrm{s} =  v_\mathrm{s}  2\pi/L $.  Hence,  
\begin{equation}  
	 \langle \hat{S}^z(r=0, \tau) \, \hat{S}^z(0,0) \rangle  \propto     
	\begin{cases}
	\frac{1}{v_\mathrm{s} \tau}    &   \text{ if  }  \;  \tau \Delta_\mathrm{s} < 1   \\
	\frac{\Delta_\mathrm{s}}{v_\mathrm{s}} e^{- \tau \Delta_\mathrm{s}} & \text{ if  } \;   \tau  \Delta_\mathrm{s}  >  1 
	\end{cases}
\, .
\end{equation}

Then, the local  spin susceptibility in the  zero-temperature  limit   and  for a finite  system  size reads  
\begin{equation}
	   \int_{0}^{\infty} d \tau  \,    \langle \hat{S}^z(r=0, \tau) \, \hat{S}^z(0,0) \rangle   \propto   \ln(L). 
\end{equation}
Since  space and time are interchangeable  for  Lorentz-invariant  systems,  similar  arguments can  be put    forward to show that  the local  
susceptibility  at  finite   temperatures  and in the  thermodynamic  limit   scales as $\ln(\beta) $.

\end{document}